\documentclass[11pt,titlepage,a4paper]{article}
\usepackage{bozhomac}  
\usepackage{bozhlogo}  
\usepackage{cite}	

\title{\bfseries	\vspace*{-1.678902345in}
{\huge On conserved operator quantities\\[1ex]
 in quantum field theory}
}

\vspace{1.7ex}

\author{
Bozhidar Z.\ Iliev
\thanks{Laboratory of Mathematical Modeling in Physics,
Institute for Nuclear Research and \mbox{Nuclear} Energy,
Bulgarian Academy of Sciences,
Boul.\ Tzarigradsko chauss\'ee~72, 1784 Sofia, Bulgaria}
\thanks{E-mail address: bozho@inrne.bas.bg}
\thanks{URL: http://theo.inrne.bas.bg/$^\sim$bozho/}
}

\date{	
 \vspace{2.27ex}\ShortTitle{Conserved operators in QFT}\\[0.27ex]
 \vspace{3.27ex}
\small
	\begin{tabular}{r@{$\colon\to~$}l}
 \vspace{0.09ex} Basic ideas	& January 5, 2003	\\[0.09ex]
 \vspace{0.09ex} Began		& January 6, 2003	\\[0.09ex]
 \vspace{0.09ex} Ended		& January 11, 2003 	\\[0.09ex]
 \vspace{0.09ex} Initial typeset& January 8 -- 14, 2003
							\\[0.09ex]
\vspace{0.09ex} Last update	& January 19, 2003	\\[0.09ex]
 \vspace{0.27ex} Produced	& \fbox{\today}	\\[0.27ex]
	\end{tabular} \\[1.27ex]
\normalsize
	\begin{tabular}{r@{$\colon~$}l}
\vspace{0.27ex} http://www.arXiv.org e-Print archive No. & hep-th/0301134
	\end{tabular} \\[-0.27ex]
 \vspace{4.27ex}{\Huge\BOZHO}	\\[4.27ex]
 \vspace{0.27ex}\Subject{Quantum field theory}
								\\[2.27ex]
	\begin{tabular}{r@{\hspace{0.512em}}|@{\hspace{0.512em}}l}
 \vspace{0.27ex}\MSC[2000]{81Q99,81T99\\\hspace{0pt}}
&
 \vspace{0.27ex}
	\end{tabular} \\[1.27ex]
 \vspace{0.27ex}\\[0.27ex]
}

\newcommand{\Hil}{\mathcal{F}}		
	\newcommand{\base}{\mathit{M}}	

 \newcommand{\dyn}[1]{\pmb{\mathbb{#1}}}	

\newcommand{\ope}[2][{}]{\lindex[\mathcal{#2}]{}{#1}} 
\begin{document}		

\renewcommand{\thepage}{\roman{page}}

\renewcommand{\thefootnote}{\fnsymbol{footnote}} 
\maketitle				
\renewcommand{\thefootnote}{\arabic{footnote}}   

\tableofcontents		


	\begin{abstract}
Conserved operator quantities in quantum field theory can be defined via
the Noether theorem in the Lagrangian formalism and as generators of some
transformations. These definitions lead to generally different conserved
operators which are suitable for different purposes. Some relations involving
conserved operators are analyzed.
	\end{abstract}

\renewcommand{\thepage}{\arabic{page}}

\section {Introduction}
\label{Introduction}

	There are two approaches for introduction of conserved operator
quantities in quantum field theory. The first one is based on the Lagrangian
formalism and defines them via the first Noether theorem as conserved
operators corresponding to smooth transformations living invariant the action
integral of an investigated system; these are the canonical conserved
operators. The second set of conserved operators consists of generators of
some transformations of state vectors (and observables). Since these operators
are of pure mathematical origin, we call them mathematical conserved
quantities (operators). The present paper is devoted to a discussion of some
relations between the mentioned two kinds of conserved quantities in quantum
field theory. It is pointed that the two types of conserved operators are
generally different and may coincide on some subspace of the system's Hilbert
space of states.

	The present work generalizes part of the results
of~\cite{bp-QFT-momentum-operator,bp-QFT-angular-momentum-operator} and may
be considered as a continuation of these papers.

	In what follows, we suppose that there is given a system of quantum
fields, described via field operators $\varphi_i(x)$,
$i=1,\dots,n\in\field[N]$, $x\in\base$ over the 4\ndash dimensional Minkowski
spacetime $\base$ endowed with standard Lorentzian metric tensor
$\eta_{\mu\nu}$ with signature $(+\,-\,-\,-)$.%
\footnote{~%
The quantum fields should be regarded as operator-valued distributions
(acting on a relevant space of test functions) in the rigorous mathematical
setting of Lagrangian quantum field theory. This approach will be considered
elsewhere.%
}
The system's Hilbert space of
states is denoted by $\Hil$ and all considerations are in Heisenberg picture
of motion if the opposite is not stated explicitly. The Greek indices
$\mu,\nu,\dots$ run from 0 to $3=\dim\base-1$ and the Einstein's summation
convention is assumed over indices repeated on different levels. The
coordinates of a point $x\in\base$ are denoted by $x^\mu$,
$\bs{x}:=(x^1,x^2,x^3)$, $\Id^3\bs{x}:=\Id x^1 \Id x^2 \Id x^3$,
and the derivative with respect to
$x^\mu$ is $\frac{\pd}{\pd x^\mu}=:\pd_\mu$. The imaginary unit is denoted
by $\iu$ and $\hbar$ and $c$ stand for the Planck's constant (divided by
$2\pi$) and the velocity of light in vacuum, respectively.


\section{Canonical conserved quantities}
	\label{Sect2}

	Suppose a system of \emph{classical} fields $\varphi_i(x)$,
$i=1,\dots,n\in\field[N]$, over the Minkowski spacetime $M$, $x\in M$, is
described via a Lagrangian $L$ depending on them and their first partial
derivatives $\pd_\mu\varphi_i(x)=\frac{\pd\varphi_i(x)}{\pd x^\mu}$,
$\{x^\mu\}$ being the (local) coordinates of $x\in M$, \ie
 $L=L(\varphi_j(x), \pd_\nu\varphi_i(x))$. Here and henceforth the Greek
indices $\mu,\nu,\dots$ run from $0$ to $\dim M-1=3$ and the Latin indices
$i,j,\dots$ run from 1 to some integer $n$. The equations of motion for
$\varphi_i(x)$, known as the \emph{Euler\ndash Lagrange equations}, are%
\footnote{~%
In this paper the Einstein's summation convention over indices appearing twice
on different levels is assumed over the whole range of their values.%
}
\(
\frac{\pd L}{\pd \varphi_i(x)}
-
\frac{\pd}{\pd x^\mu} \Bigl( \frac{\pd L}{\pd (\pd_\mu\varphi_i(x))} \Bigr)
= 0
\)
and are derived from the variational principle of stationary action, known
as the \emph{action principle} (see, e.g,~\cite[\S~1]{Bogolyubov&Shirkov},
\cite[\S~67]{Bjorken&Drell-2}, \cite[pp.~19\Ndash20]{Roman-QFT}).

	The (first) Noether theorem~\cite[\S~2]{Bogolyubov&Shirkov} says
that, if the action's variation is invariant under $C^1$ transformations
	\begin{gather}
			\label{2.2new}
	\begin{split}
& x\mapsto x^\omega = x^\omega(x)
\quad x^\omega|_{\omega=\bs0} = x
\qquad \omega=(\omega^{(1)},\dots,\omega^{(s)})
\\
& \varphi_i(x) \mapsto \varphi_i^\omega(x^\omega)
\quad \varphi_i^\omega(x^\omega)|_{\omega=\bs0} = \varphi_i(x)
	\end{split}
\\\intertext{depending on $s\in\field[N]$ independent real parameters
$\omega^{(1)},\dots,\omega^{(s)}$, then the quantities (`Noether currents')}
			\label{2.2}
\theta_{(\alpha)}^{\mu}(x)
:=
- \pi^{i\mu} \Bigl\{
\frac{\pd\varphi_i^\omega(x^\omega)} {\pd\omega^{(\alpha)}} \Big|_{\omega=0}
 -
(\pd_\nu\varphi_i(x))
\frac{\pd x^{\omega\,\nu}} {\pd\omega^{(\alpha)}} \Big|_{\omega=0}
	\Bigr\}
- L(x) \frac{\pd x^{\omega\,\mu}} {\pd\omega^{(\alpha)}} \Big|_{\omega=0} ,
\intertext{where $\alpha=1,\dots,s$ and}
			\label{2.3}
\pi^{i\mu} := \frac{\pd L}{\pd (\pd_\mu\varphi_i(x))} ,
\intertext{are conserved in a sense that their divergences vanish, \viz}
			\label{2.4}
\pd_\mu \theta_{(\alpha)}^{\mu}(x) = 0.
	\end{gather}
Respectively, the quantities
	\begin{gather}	\label{2.6}
C_{(\alpha)}(x) := \frac{1}{c} \int \theta_{(\alpha)}^0 \Id^3\bs{x} ,
\\\intertext{which in fact may depend only on $x^0$, are conserved in a sense
that}
			\label{2.7}
\frac{\pd C_{(\alpha)}(x)}{\pd x^0} = 0
	\end{gather}
and hence $\pd_\mu C_{(\alpha)}=0$. The functions (constants) $C_{(\alpha)}$
are called \emph{canonical (Noetherian, dynamical) conserved quantities}
corresponding to the symmetry transformations~\eref{2.2new} of the system
considered.

	Let us turn now our attention to a system of \emph{quantum} fields
represented by \emph{field operators} $\varphi_i(x)\colon\Hil\to\Hil$,
$i=1,\dots,n\in\field[N]$, acting on the system's Hilbert space $\Hil$ of
states and described via a Lagrangian
$\ope{L}=\ope{L}(x)=\ope{L}(\varphi_i(x),\pd_\mu\varphi_j(x))$.
Supposed the system's action integral is invariant under the  $C^1$
transformations~\eref{2.2new}. As a consequence of that supposition, one may
expect the \emph{operators}~\eref{2.2}, with $\pi^{i\mu}$ defined
via~\eref{2.3}, to be conserved, \ie the equations~\eref{2.4} to be valid.
However, at this point two problems arise: (i)~what is the meaning of the
derivatives in~\eref{2.3} as $\pd_\mu\varphi_i(x)$ is \emph{operator}, not a
classical function? and (ii)~in what order one should write the operators
compositions in~\eref{2.2}, \eg shall we write
$\pi^{i\mu}\circ\pd_\nu\varphi_i(x)$ or
$\pd_\nu\varphi_i(x)\circ\pi^{i\mu}$?
Usually~\cite{Bjorken&Drell-2,Bogolyubov&Shirkov,Itzykson&Zuber} these
problems are solved by (implicitly) adding to the theory additional
assumptions concerning the operator ordering in~\eref{2.2} and meaning of
derivatives with respect to operator\ndash valued arguments.%
\footnote{~%
E.g., derivatives like the ones in~\eref{2.3} are calculated according to the
rules of classical analysis of commuting variables by preserving the relative
order of all terms in the Lagrangian. As pointed
in~\cite{bp-QFT-action-principle}, this rule corresponds to field variations
proportional to the identity mapping $\id_\Hil$ of $\Hil$.%
}
In the work~\cite{bp-QFT-action-principle} we demonstrated that there is only
one problem connected with a suitable definition of derivatives relative to
operator\ndash valued arguments and all other results follow directly from
the (Schwinger's) action principle. The main point is that such derivatives
are mappings form (a subset of) the space $\{\Hil\to\Hil\}$ of operators on
$\Hil$ into $\{\Hil\to\Hil\}$ rather than operators $\Hil\to\Hil$. In
particular, we have
	\begin{equation}	\label{2.8}
\pi^{i\mu}(x) := \frac{\pd L}{\pd(\pd_\mu\varphi_i(x))}
\colon
\{\Hil\to\Hil\} \to \{\Hil\to\Hil\} .
	\end{equation}
For details and the rigorous definition of a derivative (of polynomial or
convergent power series) relative to operator\ndash valued argument, the
reader is referred to~\cite{bp-QFT-action-principle}. Accepting~\eref{2.8},
we can write the quantum field analogue of~\eref{2.2}, \ie the `Noether's
current operators", as
	\begin{equation}	\label{2.9}
\theta_{(\alpha)}^{\mu}(x)
:=
- \sum_{i} \pi^{i\mu}(x)
\Bigl(
\frac{\pd\varphi_i^\omega(x^\omega)} {\pd\omega^{(\alpha)}} \Big|_{\omega=0}
\Bigr)
+
\sum_{i,\nu}\pi^{i\mu}(x)
\bigl( \pd_\nu\varphi_i(x) \bigr)
\frac{\pd x^{\omega\,\nu}} {\pd\omega^{(\alpha)}} \Big|_{\omega=0}
-
L(x) \frac{\pd x^{\omega\,\mu}} {\pd\omega^{(\alpha)}} \Big|_{\omega=0} ,
	\end{equation}
which immediately leads to the conservation laws~\eref{2.4} and~\eref{2.7}.
The quantities~\eref{2.6}, with $\theta_{(\alpha)}^\mu$ given by~\eref{2.9},
are called the \emph{canonical (Noetherian, dynamical) conserved operators}
corresponding to the symmetry transformations~\eref{2.2new}.

	We end this section by the remark that the momentum, (total) angular
momentum, and charge conserved operators are generated respectively by 	the
transformation:
	\begin{subequations}	\label{2.10}
	\begin{alignat}{2}	\label{2.10a}
& x\mapsto x+b
&&
 \varphi_i(x)\mapsto \varphi_i(x)
\\			\label{2.10b}
&x^\varkappa\mapsto x^\varkappa + \varepsilon^{\varkappa\nu} x_\nu
&\quad&
\varphi_i(x)\mapsto \varphi_i(x)
  + \frac{1}{2} I_{i\mu\nu}^{j} \varepsilon^{\mu\nu} \varphi_j(x) + \dotsb
\\			\label{2.10c}
& x\mapsto x
&&
\varphi_i(x)\mapsto \e^{\frac{q}{\ih c}\lambda} \varphi_i(x) ,
	\end{alignat}
	\end{subequations}
where $b\in\base$, $\varepsilon^{\mu\nu}=-\varepsilon^{\nu\mu}\in\field[R]$,
and $\lambda\in\field[R]$ are the parameters of the corresponding
transformations, $x_\mu$ are the covariant coordinates of $x\in\base$, the
numbers $I_{i\mu\nu}^{j}=-I_{i\nu\mu}^{j}$ characterize the behaviour of the
field operators under rotations, and the dots stand for higher order terms in
$\varepsilon^{\mu\nu}$.


\section {On observer dependence of state vectors and observables}
\label{Sect3}

	Let two observers $O$ and $O'$ investigate one and the same system of
quantum fields. The quantities relative to $O'$ will be denoted as those
relative to $O$ by adding a prime to their kernel symbols. The transition
$O\mapsto O'$ implies the changes
	\begin{equation}	\label{3.1}
x\mapsto x'=L(x)
	\end{equation}
(of the coordinates) of a spacetime point $x=(x^0,x^1,x^2,x^3)\in\base$ and
	\begin{equation}	\label{3.2}
\ope{X}(x)\mapsto \ope{X}'(x')=\Lambda(\ope{X}(x))
	\end{equation}
of a state vector $\ope{X}(x)\in\Hil$ of system of quantum fields
$\varphi_i(x)$.%
\footnote{~%
It is inessential for the following whether $L$ ($\Lambda$) is an element (of
a representation) of the Poincar\'e group or not; the former case is realized
when $O$ and $O'$ are inertial observers.%
}
Requiring preservation of the scalar products in $\Hil$ under the change
$O\mapsto O'$, which physically corresponds to preservation of probability
amplitudes, we see that  $\Lambda$ is a
\emph{unitary} operator,
	\begin{equation}	\label{3.3}
\Lambda^{-1} = \Lambda^\dag
	\end{equation}
where the dagger $\dag$ denotes Hermitian conjugation (i.e., in mathematical
terms, $\Lambda^\dag$ is the adjoint to $\Lambda$ operator).

	 Let $\dyn{A}$ be a dynamical variable and
$\ope{A}(x)\colon\Hil\to\Hil$ be the corresponding to it observable. The
change $O\mapsto O'$ entails $\ope{A}(x)\mapsto \ope{A}(L(x))$. Supposing
preservation of the mean (expectation) values (and the matrix elements of
$\dyn{A}$ (or $\ope{A}(x)$)) in states with finite norm under the change
$O\mapsto O'$, we get
	\begin{equation}	\label{3.4}
\ope{A}(L(x))
= (\Lambda^\dag)^{-1} \circ \ope{A}(x) \circ \Lambda^{-1}
= \Lambda \circ \ope{A}(x) \circ \Lambda^{-1}.
	\end{equation}
As explained in~\cite[sect.~4]{bp-QFT-angular-momentum-operator} or
in~\cite{Bjorken&Drell-2,Bogolyubov&Shirkov}, the field operators
$\varphi_i(x)$ undergo more complicated change when one passes from $O$ to
$O'$:
	\begin{equation}	\label{3.5}
\varphi_i(x) \mapsto \sum_{j} \Sbrindex[(S^{-1})]{i}{j}(L) \varphi_j(x)
=
\Lambda\circ \varphi_i(x) \circ \Lambda^{-1}
	\end{equation}
where the depending on $L$ matrix $S=S(L)=[\Sbrindex[(S^{-1})]{i}{j}(L)]$
characterizes the transformation properties of any particular field (\eg
scale or vector one) under $O\mapsto O'$ and is such that
$S(L)|_{L=\id_{\base}}$ is the identity matrix of relevant size.


\section {Transformations with Hermitian generators}
\label{Sect4}

	Let $\omega^1,\dots,\omega^s$, $s\in\field[N]$, be real independent
parameters and $\omega:=(\omega^1,\dots,\omega^s)\in\field[R]^s$. Suppose the
changes~\eref{3.1} and~\eref{3.2} depend on $\omega$ and
	\begin{equation}	\label{4.1}
	\begin{split}
& x\mapsto x' = L^\omega(x) = x^\omega(x) \quad x^\omega(x)|_{\omega=0} = x
\\
& \Lambda
 = \Lambda^\omega
 = \exp\Bigl\{ \
   \frac{\eta}{\ih} \sum_{\alpha=1}^{s} \omega^\alpha
					\ope{J}_\alpha^{\mathrm{m}}
   \Bigr\} ,
	\end{split}
	\end{equation}
where the operators $\ope{J}_\alpha^{\mathrm{m}}\colon\Hil\to\Hil$
are Hermitian,
	\begin{equation}	\label{4.2}
(\ope{J}_\alpha^{\mathrm{m}})^\dag = \ope{J}_\alpha^{\mathrm{m}} ,
	\end{equation}
which ensures the validity of~\eref{3.3}, and the particular choice of the
constant $\eta\in\field[R]\setminus\{0\}$ depends on what physical
interpretation of $\ope{J}_\alpha^{\mathrm{m}}$ one intends to get.

	Differentiating~\eref{3.4} and~\eref{3.5} with respect to
$\omega^\alpha$ and setting $\omega=0$, we rewrite them in differential form
respectively as
	\begin{align}	\label{4.3}
&
\eta [\ope{A}(x) , \ope{J}_\alpha^{\mathrm{m}}]_{\_}
=
- \ih \frac{\pd\ope{A}(x)}{\pd x^\mu}
    \frac{\pd x^{\omega\,\mu}}{\pd \omega^\alpha}\Big|_{\omega=0}
\\			\label{4.4}
&
\eta [\varphi_i(x) , \ope{J}_\alpha^{\mathrm{m}}]_{\_}
=
\ih \sum_{j} I_{i\alpha}^{j} \varphi_j(x)
- \ih \frac{\pd\varphi_i(x)}{\pd x^\mu}
    \frac{\pd x^{\omega\,\mu}}{\pd \omega^\alpha}\Big|_{\omega=0}
	\end{align}
where
\(
I_{i\alpha}^{j}
:= \frac{\pd \Sbrindex[S]{i}{j}(L^\omega)}{\pd \omega^\alpha} \Big|_{\omega=0}
\),
\ie
\(
\Sbrindex[S]{i}{j}(L^\omega)
= \delta_i^j + \sum_{\alpha} I_{i\alpha}^{j} \omega^\alpha + \dotsb
\)
with $\delta_i^j$ being the Kroneker deltas and the dots denoting higher
order terms in $\omega$,
and $[\ope{A},\ope{B}]_{\_}:=\ope{A}\circ \ope{B} - \ope{B}\circ \ope{A}$ is
the commutator of operators $\ope{A},\ope{B}\colon\Hil\to\Hil$.

	In particular, to describe the quantum analogue of the
transformations~\eref{2.10}, in~\eref{4.1} we have to make respectively the
replacements:
	\begin{subequations}	\label{4.5}
	\begin{alignat}{4}	\label{4.5a}
& \omega^\alpha\mapsto b^\mu
&\quad& x^\omega\mapsto x^b=x+b
&\quad& \eta\mapsto -1
&\quad& \ope{J}_\alpha^{\mathrm{m}}\mapsto \ope{P}_{\mu}^{\mathrm{t}}
\\			\label{4.5b}
& \omega^\alpha\mapsto \varepsilon^{\mu\nu} \quad (\mu<\nu)
&& x^{\omega\,\mu} \mapsto
		x^{\varepsilon\,\mu} = x^\mu + \varepsilon^{\mu\nu} x_\nu
&& \eta\mapsto +1
&& \ope{J}_\alpha^{\mathrm{m}}\mapsto \ope{M}_{\mu\nu}^{\mathrm{r}}
\\			\label{4.5c}
& \omega^\alpha\mapsto \lambda
&& x^{\omega} \mapsto x^\lambda = x
&& \eta\mapsto \frac{q}{c}
&& \ope{J}_\alpha^{\mathrm{m}}\mapsto \ope{Q}^{\mathrm{p}} ,
	\end{alignat}
	\end{subequations}
so that
$\frac{\pd x^{\omega\,\varkappa}}{\pd \omega^\alpha}\big|_{\omega=0}$
reduces to
 $\delta_\mu^\varkappa$,
 $(\delta_\mu^\varkappa x_\nu - \delta_\nu^\varkappa x_\mu $), and
 $0\in\field[R]$, respectively.
The operators $\ope{P}_{\mu}^{\mathrm{t}}$, $\ope{M}_{\mu\nu}^{\mathrm{r}}$,
and $\ope{Q}^{\mathrm{p}}$ are the translation (mathematical) momentum
operator, total rotational (mathematical) angular momentum operator, and
constant phase transformation (mathematical) charge operator, respectively.
In these cases, the equations~\eref{4.4} are known as the Heisenberg
equations/relations for the operators
mentioned~\cite{Roman-QFT,Bogolyubov&Shirkov,Bjorken&Drell-2} . For that
reason, it is convenient to call~\eref{4.4}
\emph{Heisenberg equations/relations} (for the operators
$\ope{J}_\alpha^{\mathrm{m}}$) in the general case.

	The transformations~\eref{3.1} and~\eref{3.5}, defined by the
choice~\eref{4.1}, are the \emph{quantum} observer\ndash transformation
version of~\eref{2.2new}. For that reason, one can expect the (spacetime
constant) operators $\ope{J}_\alpha^{\mathrm{m}}$ to play, in some sense, a
role similar to the conserved operators~\eref{2.9}; we shall call
$\ope{J}_\alpha^{\mathrm{m}}$ \emph{mathematical conserved operators}
corresponding to the transformations~\eref{3.1} and~\eref{3.5} under the
choices~\eref{4.1}.

	Suppose there exist operators $\ope{J}_\alpha^{\mathrm{QM}}$, where
QM stands for quantum mechanics%
\footnote{~%
This notation reminds only some analogy with quantum mechanics. If one
identifies $\Hil$ with the Hilbert space of this theory and makes some other
assumptions, (part of) the generators $\ope{J}_\alpha^{\mathrm{QM}}$ will
coincide with similar objects in quantum mechanics. However, as the Hilbert
spaces of quantum field theory and quantum mechanics are different, the
corresponding operators in these theories cannot be identified. See similar
remarks in~\cite{bp-QFT-momentum-operator,bp-QFT-angular-momentum-operator}
concerning the momentum and angular momentum operators, respectively.%
},
generating the change $\ope{X}(x)\mapsto\ope{X}(x')$, \ie such that ($|\eta|$
is the absolute value of $\eta$)
	\begin{equation}	\label{4.6}
\ope{X}(x)\mapsto\ope{X}(x')
= \Lambda^{\mathrm{QM}} (\ope{X}(x))
:=
\exp\Bigl\{
\frac{|\eta|}{\ih} \sum_{\alpha=1}^{s} \omega^\alpha
						\ope{J}_\alpha^{\mathrm{QM}}
\Bigr\}
\bigl(\ope{X}(x)\bigr) .
	\end{equation}
Note that $\ope{J}_\alpha^{\mathrm{QM}}$ (as well as
$\ope{J}_\alpha^{\mathrm{m}}$) may depend on $x$; for instance, the changes
$x\mapsto x^\omega$ defined via~\eref{4.5a}--\eref{4.5c} entail~\eref{4.6}
with respectively ($\id_\Hil$ is the identity mapping of $\Hil$)
	\begin{subequations}	\label{4.7}
	\begin{align}	\label{4.7a}
&
\ope{J}_\alpha^{\mathrm{QM}}\mapsto
   \ope{P}_{\mu}^{\mathrm{QM}} = \ih \pd_\mu
\\			\label{4.7b}
&
\ope{J}_\alpha^{\mathrm{QM}}\mapsto
   \ope{M}_{\mu\nu}^{\mathrm{QM}} = \ih (x_\mu\pd_\nu - x_\nu\pd_\mu)
\\			\label{4.7c}
&
\ope{J}_\alpha^{\mathrm{QM}}\mapsto
    \ope{Q}_{\mu}^{\mathrm{QM}} = \e^{\frac{q}{\ih c}\lambda} \id_\Hil .
	\end{align}
	\end{subequations}

	The transformation~\eref{4.6} implies the changes
	\begin{align}	\label{4.8}
\ope{A}(x) \mapsto \ope{A}(x')
= \Lambda^{\mathrm{QM}} \circ \ope{A}(x) \circ (\Lambda^{\mathrm{QM}})^{-1}.
\\			\label{4.9}
\varphi_i(x) \mapsto \sum_{j} \Sbrindex[(S^{-1})]{i}{j}(L) \varphi_j(x')
=
\Lambda^{\mathrm{QM}}\circ \varphi_i(x) \circ (\Lambda^{\mathrm{QM}})^{-1}
	\end{align}
which, in differential form, entail
	\begin{align}	\label{4.10}
&
|\eta| [\ope{A}(x) , \ope{J}_\alpha^{\mathrm{QM}}]_{\_}
=
- \ih \frac{\pd\ope{A}(x)}{\pd x^\mu}
    \frac{\pd x^{\omega\,\mu}}{\pd \omega^\alpha}\Big|_{\omega=0}
\\			\label{4.11}
&
|\eta| [\varphi_i(x) , \ope{J}_\alpha^{\mathrm{QM}}]_{\_}
=
  \ih \sum_{j} I_{i\alpha}^{j} \varphi_j(x)
- \ih \frac{\pd\varphi_i(x)}{\pd x^\mu}
    \frac{\pd x^{\omega\,\mu}}{\pd \omega^\alpha}\Big|_{\omega=0}
	\end{align}
Comparing these equations with~\eref{4.3} and~\eref{4.4}, we find
	\begin{subequations}	\label{4.12}
	\begin{align}	\label{4.12a}
&
[ \ope{A}(x) ,
\ope{J}_\alpha^{\mathrm{m}} - \sign \eta \ope{J}_\alpha^{\mathrm{QM}} ]_{\_}=0
\\			\label{4.12b}
&
[ \varphi_i(x) ,
\ope{J}_\alpha^{\mathrm{m}} - \sign \eta \ope{J}_\alpha^{\mathrm{QM}} ]_{\_}=0
	\end{align}
	\end{subequations}
where  $\sign \eta:=\eta/|\eta|\in\{-1.+1\}$ is the sign of
$\eta\in\field[R]\setminus\{0\}$. If we admit~\eref{4.12a} to hold for
\emph{every} $\ope{A}(x)\colon\Hil\to\Hil$, the Schur's lemma%
\footnote{~%
See, e.g,~\cite[appendix~II]{Rumer&Fet}, \cite[sec.~8.2]{Kirillov-1976},
\cite[ch.~5, sec.~3]{Barut&Roczka}.%
}
implies
	\begin{equation}	\label{4.13}
\ope{J}_\alpha^{\mathrm{m}}
= \sign \eta \ope{J}_\alpha^{\mathrm{QM}} + j_\alpha\id_\Hil,
	\end{equation}
where $j_\alpha$ are real numbers (with the same dimension as the eigenvalues
of $\ope{J}_\alpha^{\mathrm{m}}$).


\section {Discussion}
\label{Sect5}

	Following the opinion established in the literature%
\footnote{~%
See also the
papers~\cite{bp-QFT-momentum-operator,bp-QFT-angular-momentum-operator} in
which the momentum and angular momentum are analyzed.%
},
the identification
	\begin{equation}	\label{5.1}
C_{(\alpha)} = \ope{J}_\alpha^{\mathrm{m}}
	\end{equation}
may seem `natural' \emph{prima facie} but, generally, it is unacceptable as
its l.h.s.\ comes out from the Lagrangian formalism (via~\eref{2.9}
and~\eref{2.6}), while its r.h.s.\ originates from pure mathematical
(geometrical) considerations and is suitable for the axiomatic quantum field
theory~\cite{Bogolyubov&et_al.-AxQFT,Bogolyubov&et_al.-QFT}.

	As an equality weaker than~\eref{5.1}, the Heisenberg
relations~\eref{4.4} with $C_{(\alpha)}$ for $\ope{J}_\alpha^{\mathrm{m}}$
can be assumed:
	\begin{equation}	\label{5.2}
\eta [\varphi_i(x) , C_{(\alpha)}]_{\_}
=
\ih \sum_{j} I_{i\alpha}^{j} \varphi_j(x)
- \ih \frac{\pd\varphi_i(x)}{\pd x^\mu}
    \frac{\pd x^{\omega\,\mu}}{\pd \omega^\alpha}\Big|_{\omega=0} .
	\end{equation}
However, these equations as well as~\eref{5.1} are external to the Lagrangian
formalism by means of which the canonical conserved operators are defined. As
discussed in~\cite[\S~68]{Bjorken&Drell-2} on particular examples, the
validity of the equations~\eref{5.2} should be checked for any particular
Lagrangian and they express (in the sense explained in \emph{loc.\ cit.})\ the
relativistic covariance of the Lagrangian quantum field theory.

	Generally the equation~\eref{4.3} with $C_{(\alpha)}$ for
$\ope{J}_\alpha^{\mathrm{m}}$, viz.
	\begin{equation}	\label{5.3}
\eta [\ope{A}(x) , C_{(\alpha)}]_{\_}
=
- \ih \frac{\pd\ope{A}(x)}{\pd x^\mu}
    \frac{\pd x^{\omega\,\mu}}{\pd \omega^\alpha}\Big|_{\omega=0} ,
	\end{equation}
cannot hold; a counterexample being the choice of $\ope{A}(x)$ and
$C_{(\alpha)}$ as the momentum and angular momentum operators (or \emph{vice
versa}). If~\eref{5.3} happens to be valid for operators $\ope{A}(x)$ forming
an irreducible representation of some group, then, by virtue of~\eref{5.3}
and~\eref{4.3}, the Schur's lemma implies
	\begin{equation}	\label{5.4}
C_{(\alpha)}
= \sign\eta \ope{J}_\alpha^{\mathrm{m}} + i_\alpha \id_\Hil
= \sign\eta \ope{J}_\alpha^{\mathrm{QM}} + (i_\alpha + j_\alpha) \id_\Hil
	\end{equation}
for some real numbers $i_\alpha$ (see also~\eref{4.13}).

	Let a vector $\ope{X}\in\Hil$ represents a state of the system of
quantum fields considered. It is a spacetime\ndash constant vector as we are
working in Heisenberg picture of motion. Consequently, we have
$\ope{X}(x)=\ope{X}(x')$ which, when combined with~\eref{4.6}, entails
	\begin{equation}	\label{5.5}
\ope{J}_\alpha^{\mathrm{QM}} (\ope{X}) = 0.
	\end{equation}
So, applying~\eref{4.13} to $\ope{X}$, we get
	\begin{equation}	\label{5.6}
\ope{J}_\alpha^{\mathrm{m}} (\ope{X}) = j_\alpha \ope{X} .
	\end{equation}
If one intends to interpret $\ope{J}_\alpha^{\mathrm{m}}$ as the conserved
canonical operators $C_{(\alpha)}$ (see the possible equality~\eref{5.1}),
then one should interpret $j_\alpha$ as the mean (expectation) value of
$C_{(\alpha)}$, which will be the case if
	\begin{equation}	\label{5.7}
C_{(\alpha)} (\ope{X}) = j_\alpha \ope{X} .
	\end{equation}
(Notice,~\eref{5.7} and~\eref{5.4} are compatible iff $i_\alpha=0$.) The
equations~\eref{5.6} and~\eref{5.7} imply
	\begin{equation}	\label{5.8}
C_{(\alpha)}|_{\ope{D}_j}  = \ope{J}_\alpha^{\mathrm{m}}|_{\ope{D}_j} .
	\end{equation}
where
	\begin{equation}	\label{5.9}
\ope{D}_j
:= \{ \ope{X}\in\Hil : C_{(\alpha)}(\ope{X}) = j_\alpha \ope{X} \}.
	\end{equation}
Generally the set $\ope{D}_j$ is a proper subset of $\Hil$ and
hence~\eref{5.8} is weaker than~\eref{5.1}; if a basis of $\Hil$ can be
formed from vectors in $\ope{D}_j$, then~\eref{5.8} and~\eref{5.1} will be
equivalent. But, in the general case, equations~\eref{5.2} and~\eref{4.4}
lead only to
	\begin{equation}	\label{5.10}
[ \varphi_i(x) , C_{(\alpha)} ]_{\_}
=
[ \varphi_i(x) , \ope{J}_\alpha^{\mathrm{m}} ]_{\_}
\qquad
\Bigl(=
\frac{1}{\eta}
\ih \sum_{j} I_{i\alpha}^{j} \varphi_j(x)
-
\frac{1}{\eta}
\ih \frac{\pd\varphi_i(x)}{\pd x^\mu}
    \frac{\pd x^{\omega\,\mu}}{\pd \omega^\alpha}\Big|_{\omega=0}
\Bigr) ,
	\end{equation}
but not to~\eref{5.1}.

	Ending this section, we note that the equality
$C_{(\alpha)}=\ope{J}_\alpha^{\mathrm{QM}}$ is unacceptable as, in view
of~\eref{5.5}, it leads to identically vanishing eigenvalues of
$C_{(\alpha)}$.


\section {Conclusion}
\label{Conclusion}

	In this work we have analyzed two types of conserved operator
quantities in quantum field theory, \viz the ones arising from the (first)
Noether theorem in the framework of Lagrangian formalism and conserved
operators having pure mathematical origin as generators of some
transformations (and having natural place in the axiomatic approach). These
operators are generally different and their equality is a problem which is
external to the Lagrangian formalism and may be considered as possible
subsidiary restrictions to it. However, using the arbitrariness~\eref{4.13}
in the mathematical conserved operators, both types of conserved operators
can be chosen to coincide on the set~\eref{5.9}. As weaker conditions
additionally imposed on the Lagrangian formalism, one can require the
equality~\eref{5.10} between the commutators of the field operators and
conserved operators. As it is known~\cite{Bjorken&Drell-2}, the Heisenberg
relations~\eref{5.2} are equations relative to the field operators,
while~\eref{4.4} are identities with respect to them.


\addcontentsline{toc}{section}{References}
\bibliography{bozhopub,bozhoref}
\bibliographystyle{unsrt}
\addcontentsline{toc}{subsubsection}{This article ends at page}

\end{document}